\documentclass{article}
\usepackage{spconf,amsmath,graphicx}
\usepackage{amsfonts}
\usepackage{amsmath}
\usepackage{adjustbox}
\usepackage{bbm}

\usepackage{algorithmicx}
\usepackage{algpseudocode}
\usepackage[ruled]{algorithm}
\usepackage{graphicx}
\usepackage{caption}
\usepackage{graphicx,times,amsmath} % Add all your packages here
\usepackage{caption}
\usepackage{subcaption}
\usepackage[rightcaption]{sidecap}
\usepackage{caption}
\usepackage{multirow}
\usepackage{hhline}
\usepackage{amsmath}
\usepackage{epstopdf}
\usepackage{tabularx}
\usepackage{tikz}
\usetikzlibrary{matrix}
\usepackage{mathtools, nccmath}
\usepackage{pst-node}

\graphicspath{ {images/} }

\usepackage{alphalph}

\title{Joint Human Orientation-Activity Recognition Using WiFi Signals for\\ Human-Machine Interaction }

\usepackage{fancyhdr}
\name{\begin{tabular}{c}Hojjat Salehinejad$^{1,2}$, Member, IEEE, Navid Hasanzadeh$^{2}$,Radomir Djogo$^{2}$, and \\
\textit{Shahrokh Valaee}$^{2}$, \textit{Fellow, IEEE}\end{tabular}
\vspace{-4mm}
}
\address{$^1$Kern Center for the Science of Health Care Delivery, Mayo Clinic, Rochester, MN, USA \\
$^2$Department of Electrical \& Computer Engineering, University of Toronto, Toronto, Canada \\
\textit{hojjat@ieee.org, \{navid.hasanzadeh, radomir.djogo\}@mail.utoronto.ca, valaee@ece.utoronto.ca}}

\begin{document}
\newcommand*{\img}{
  \includegraphics[
    width=\linewidth,
    height=20pt,
    keepaspectratio=false,
  ]{example-image-a}}

\maketitle

\begin{abstract}
WiFi sensing is an important part of the new WiFi 802.11bf standard, which can detect motion and measure distances. In recent years, some machine learning methods have been proposed for human activity recognition from WiFi signals. However, to the best of our knowledge, none of these methods have explored orientation prediction of the user using WiFi signals. Orientation prediction is particularly critical for human-machine interaction in an environment with multiple smart devices. In this paper, we propose a data collection setup and machine learning models for joint human orientation and activity recognition using WiFi signals from a single access point (AP) or multiple APs. The results show feasibility of joint orientation-activity recognition in an indoor environment with a high accuracy. 

\end{abstract}
\section{Introduction}
\label{sec:intro}
Human activity recognition (HAR) refers to detection and recognition of human gestures and activities in an environment. Some major systems/mediums for collecting data are wearable sensors (e.g. gyroscope and accelerometer), cameras (e.g. still image and video), and radio frequency signals (e.g. WiFi signals)~\cite{gu2021survey}. HAR with wireless signals has attracted attention due to its privacy preserving nature, broad sensing coverage, and ability to sense the environment without line-of-sight (LoS)~\cite{salehinejad2022litehar}. This is particularly interesting since the WiFi 802.11bf standard will enable remote monitoring and sensing~\cite{restuccia2021ieee}.

{\it Channel state information} (CSI) in a wireless communication system can provide properties about the wireless channel and how a subcarrier has been affected in the environment. Changes in the environment such as walking, falling, and sitting can affect the CSI signals which can be used for various sensing applications. CSI is measured in the baseband and is a vector of complex values. A multiple-input multiple-output (MIMO) wireless system provides a spatial diversity which can be used for wider and more accurate sensing and detection of activities.  This property of wireless signals can be very useful in designing systems for human-machine interaction. Some examples are presence detection~\cite{di2018wifi}, security systems~\cite{zhang2020wifi}, localization~\cite{salehinejad20143d}, and internet of things~\cite{khan2020differential}.  
 
 Various approaches have been proposed for HAR using machine learning. That includes random forest (RF)~\cite{yousefi2017survey}, hidden Markov model (HMM)~\cite{yousefi2017survey}, long-short-term memory (LSTM)~\cite{yousefi2017survey}, sparse
auto-encoder (SAE) network~\cite{gao2017csi}, attention-based bi-directional LSTM~\cite{chen2018wifi}, and diversified deep ensemble learning (WiARes)~\cite{cui2021device}.
Most of the proposed methods are based on training many trainable parameters for feature extraction from CSI measurements. This approach requires large CSI training data and hyper-parameter tuning. In addition, most of these models due to their high computational complexity may not be suitable for implementation on resource-limited devices such as smart phones and edge devices~\cite{salehinejad2021edropout}. LiteHAR~\cite{salehinejad2022litehar} method uses a large number of random convolution kernels without training them~\cite{dempster2020rocket} for feature extraction, followed by a pool of Ridge regression classifiers per frequency for activity recognition. This approach enables fast and accurate HAR using CSI.

To the best of our knowledge, none of the previous works in HAR have explored the possibility of predicting both activity and orientation of the user using CSI. In this paper, machine learning models for prediction of the joint user activity and orientation are introduced. Orientation prediction is particularly important for interaction with devices in smart environments, where multiple devices exist. It governs which device the user is trying to interact with. We have built an infrastructure for CSI measurements collection from multiple access points (APs). Based on our previous work for a light-weight HAR~\cite{salehinejad2022litehar} solution, the idea of using 1-dimensional random convolution kernels in~\cite{dempster2020minirocket} is utilized for feature extraction from CSI measurements. Then, Ridge regression classifiers are used for prediction of the activation and orientation of the user. The proposed models are evaluated for single AP and multiple AP scenarios and the performance results are discussed.

\section{Joint Orientation-Activity Recognition Model}

In this section, we discuss the proposed model for joint human orientation-activity recognition in an indoor environment equipped with one/multiple APs for a single user. First, the feature extraction procedure is introduced. Then, three features classification approaches are proposed. 

Let $\mathbf{X}_{a}\in\mathbb{R}_{\geq 0}^{S\times T}$ represent the CSI amplitudes of AP $a$ with $S$ subcarriers over $T$ indices (i.e. the length of CSI input). For the AP $a$, the set of $N$ CSI samples is $\{(\mathbf{X}_{a,1},c_{1},o_1),...,(\mathbf{X}_{a,N},c_{N},o_N)\}$ where $N$ is the number of samples, $c_n$ is the activity class, and $o_n$ is the orientation class for sample $n\in\{1,...,N\}$. In general, the possible orientation and activity classes are finite discrete sets. The set of activity classes is $\mathbf{c}=(c_1,...,c_C)$ and the set of orientation classes is $\mathbf{o}=(o_1,...,o_O)$, where $C$ is the number of activity classes and $O$ is the number of orientations. The set of samples can be extended for $A$ APs as $\{(\mathbf{X}_{1,1},...,\mathbf{X}_{A,1},c_{1},o_1),..., \allowbreak(\mathbf{X}_{1 ,N},...,\mathbf{X}_{A,N},c_{N},o_N)\}$.

\begin{figure}[!t]
\centering
\captionsetup{font=footnotesize}
\begin{subfigure}[t]{0.5\textwidth}
\captionsetup{font=footnotesize}
\centering
\includegraphics[width=1\textwidth]{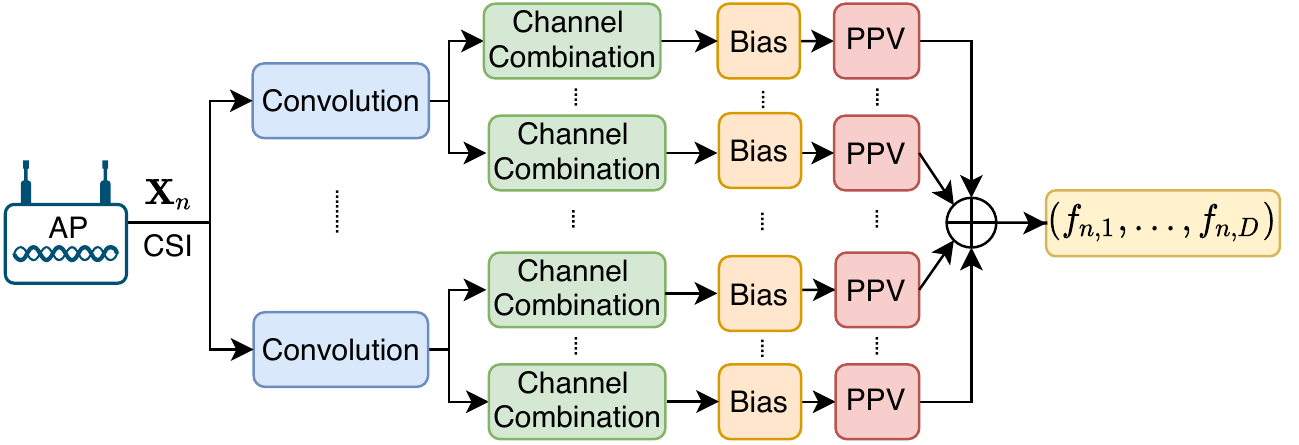} 
\caption{Feature extraction and concatenation ($\bigoplus$) for the input channel state information (CSI) $\mathbf{X}_n$. PPV refers to calculating the portion of positive values using (\ref{eq:ppv}). $D$ is the number of extracted features.}
\label{fig:feature_extract}
\end{subfigure}% 
% \vspace{4mm}

\begin{subfigure}[t]{0.44\textwidth}
\captionsetup{font=footnotesize}
\centering
\includegraphics[width=1\textwidth]{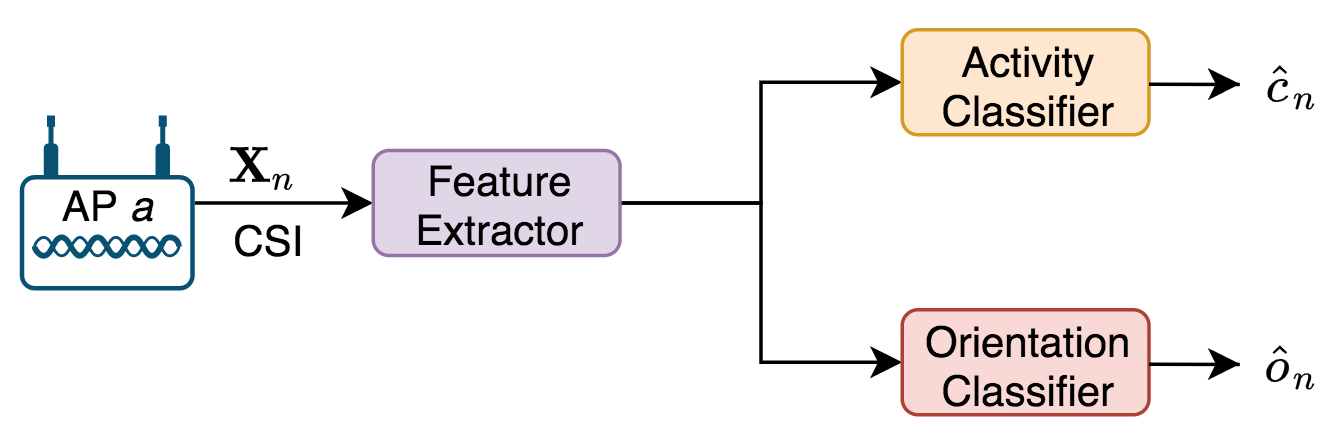}
\caption{Single access point (SAP) model using the feature extractor in (a).}
\label{fig:}
\end{subfigure}%  
% \vspace{4mm}

\begin{subfigure}[t]{0.44\textwidth}
\captionsetup{font=footnotesize}
\centering
\includegraphics[width=1\textwidth]{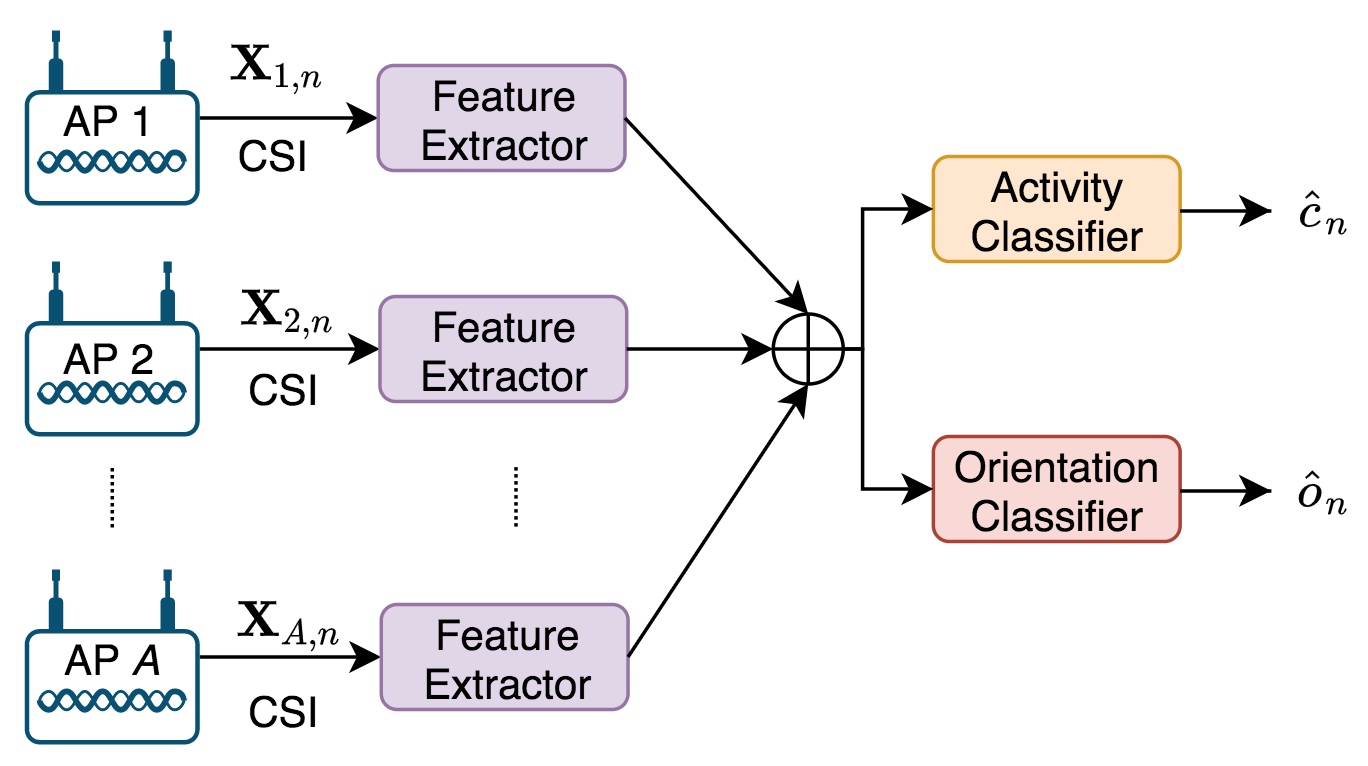}
\caption{Concatenation of multiple access points (CMAP) model using the feature extractor in (a).}
\label{fig:}
\end{subfigure}%  
% \vspace{4mm}

\begin{subfigure}[t]{0.44\textwidth}
\captionsetup{font=footnotesize}
\centering
\includegraphics[width=1\textwidth]{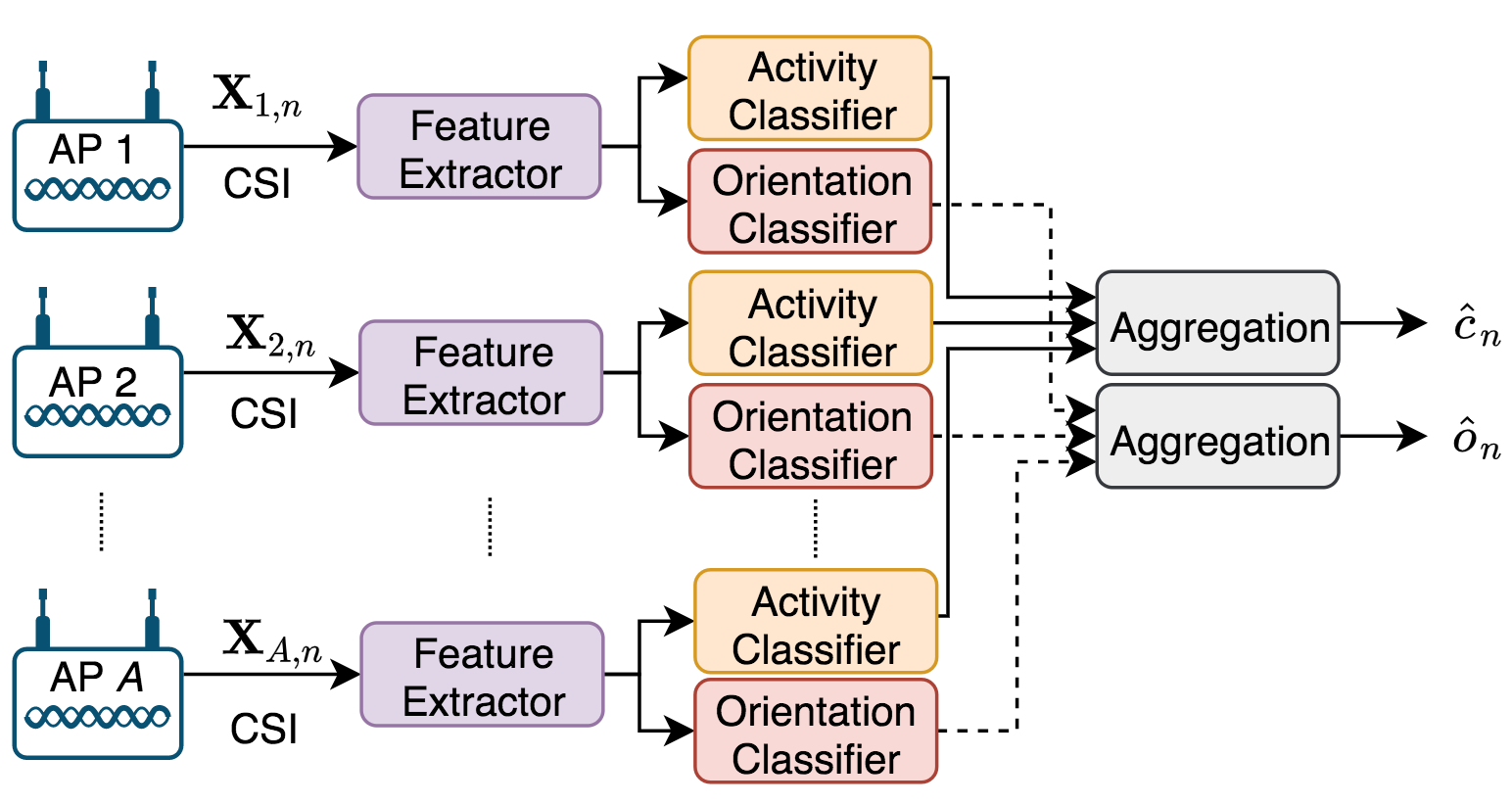}
\caption{Aggregation of multiple access points (AMAP) model using the feature extractor in (a).}
\label{fig:}
\end{subfigure}%  
\caption{Proposed feature extraction and joint orientation-activity classification models using a single access point (AP) and multiple APs.} 
\label{fig:models}
\vspace{-4mm}
\end{figure}

\subsection{Feature Extraction}
\label{sec:featureextraction}
Figure~\ref{fig:models}(a) shows the feature extraction procedure from a CSI sample $\mathbf{X}_n$ for a single AP. In this approach, based on the multivariate MiniRocket feature extraction method proposed in~\cite{dempster2021minirocket}, $K$ 1-dimensional convolution kernels $(\mathbf{w}_1,...,\mathbf{w}_K)$ are generated where the length of each kernel is fixed and the weights are selected randomly from $\{-1,2\}$. For each kernel, a set of dilation factors is generated which controls the spread of the kernel over an input with fixed length of $T$. The set of dilations for kernel $k$ is selected from ${\mathcal{L}=\{\lfloor2^{i\cdot L_{max}/L'}\rfloor|i\in(0,...,L')\}}$ where $L'$ is a constant, $L_{max}=log_2\big((T-1)/(|\mathbf{w}_k|-1)\big)$ and $L=|\mathcal{L}|$ is the cardinality of $\mathcal{L}$. This provides $K\times L$ different combinations of kernels and dilations as ${\{\mathbf{w}_{k,l}|k\in(1,...,K),l\in(1,...,L)\}}$. 
The convolution of an input CSI $\mathbf{X}$ with each kernel is 
\begin{equation}
    \mathbf{u}_{s,k,l} = \mathbf{x}_{s}*\mathbf{w}_{k,l},
\end{equation}
for $s\in(1,...,S)$, $k\in(1,...,K)$, and $l\in(1,...,L)$. 

A set of bias terms $\{b_{k,l,j}|j\in(1,...,J)\}$ is then calculated based on the quantiles of the convolution output for each pair of kernel and dilation $(k,l)$. The channel-wise features along with the bias term are then combined as
\begin{equation}
    \mathbf{v}_{k,l,j}=\sum_{s=1}^{S} \mathbf{u}_{s,k,l} - b_{k,l,j}.
    \label{eq:bias}
\end{equation}
The process of selecting the dilation and bias values is deeply discussed in~\cite{dempster2021minirocket}. It is suggested that the total number of extracted features should be kept constant (i.e. $D=9,996$) as a multiple of $K$. A feature selection method is proposed in~\cite{salehinejad2022s} for reducing $D$. The features are extracted by computing the proportion of positive values (\textit{ppv}) as
\begin{equation}
    f_{k,l,j}=\frac{1}{|\mathbf{v}_{k,l,j}|}\sum_{i=1}^{|\mathbf{v}_{k,l,j}|}\mathbbm{1}[v_{k,l,j,i}>0],
    \label{eq:ppv}
\end{equation}
for $k\in(1,...,K)$, $l\in(1,...,L)$, and $j\in(1,...,J_{k,l})$ where $J_{k,l}$ is the number of bias terms and $\mathbbm{1}[\cdot]$ is the indicator function. The features can be vectorized for the input CSI signal $\mathbf{X}_n$ as $\mathbf{f}_n=(f_{n,1},...,f_{n,D})$.

\subsection{Joint Orientation-Activity Classification}
Generally, CSI signals are collected from multiple APs in an indoor environment for HAR applications. In this section, first we introduce an approach for joint orientation-activity recognition from a single AP (SAP) based on the feature extraction procedure discussed in Subsection~\ref{sec:featureextraction}. Then, this approach is extended to introduce approaches for aggregation of extracted features from multiple APs (AMAP) and a concatenation of multiple APs (CMAP).

\begin{figure}[!t]
\centering
\captionsetup{font=footnotesize}
\centering
\includegraphics[width=0.4\textwidth]{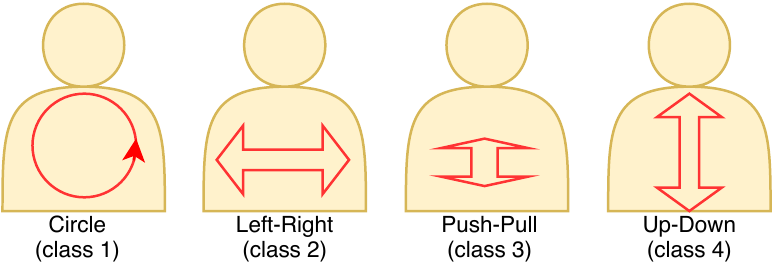} 
\caption{Four human activities (gestures) used for experiments.} 
\label{fig:activities}
\vspace{-4mm}
\end{figure}

\subsubsection{Single Access Point (SAP)}
Figure~\ref{fig:models}(b) shows the setup with a single AP for join orientation-activity recognition. For a given training dataset, the features $\mathbf{f}_n$ are extracted and passed to two Ridge regression classifiers $\hat{o}_n=\psi_{o}(\mathbf{f}_n)$ and $\hat{c}_n=\psi_{c}(\mathbf{f}_n)$, where $\hat{o}_n$ and $\hat{c}_n$ are the predicted orientation class and activation class, respectively, for the input $\mathbf{X}_n$. This is a general framework and other classifier may be used and evaluated.

\subsubsection{Concatenation of Multiple Access Points (CMAP)}
A CSI collection setup with multiple APs increases diversity of the signal collection, which enhances sensing of environment. Figure~\ref{fig:models}(c) shows a setup where $A$ APs are utilized for CSI collection and a feature extractor is implemented per AP. The extracted features are then concatenated as ${\mathbf{f}_n=\big(f_{a,n,d}|a\in(1,...,A),d\in(1,...,D)\big)}$ for each sample. The set of features $\{\mathbf{f}_n|n\in(1,...,N)\}$ and the corresponding target classes are then used for training the activity and orientation Ridge regression classifiers.

\subsubsection{Aggregation of Multiple Access Points (AMAP)}
In the AMAP approach, a feature extractor is allocated per AP followed by a dedicated activity classifier $\hat{c}_{a,n}=\phi_{a}(\mathbf{f}_{a,n})$ and orientation classifier $\hat{o}_{a,n}=\psi_{a}(\mathbf{f}_{a,n})$ for $a\in(1,...,A)$ and $n\in(1,...,N)$. For a given input $\mathbf{X}_n$, the set of predicted orientations  is $\hat{\mathbf{o}}_n=\big(\hat{o}_{a,n}|a\in(1,...,A)\big)$ and the set of predicted activities is $\hat{\mathbf{c}}_n=\big(\hat{c}_{a,n}|a\in(1,...,A)\big)$. Using an aggregation (voting) approach, the predicted activity is 
\begin{equation}
    \hat{c}_n=\underset{c\in\mathbf{c}}{\text{argmax}}(\sum_{a=1}^{A}\mathbbm{1}{[\hat{c}_{a,n},c_n]}\;|\; c_n\in\mathbf{c}),
    \label{eq:voting_activ}
\end{equation}
and the predicted orientation is 
\begin{equation}
    \hat{o}_n=\underset{o\in\mathbf{o}}{\text{argmax}}(\sum_{a=1}^{A}\mathbbm{1}{[\hat{o}_{a,n},o_n]}\;|\; o_n\in\mathbf{o}),
    \label{eq:voting_ori}
\end{equation}
where $\mathbbm{1}{[\hat{i},i]}=1$ if $\hat{i}=i$ and $\mathbbm{1}{[\hat{i},i]}=0$ otherwise.

\begin{figure}[!t]
\centering
\captionsetup{font=footnotesize}
\centering
\includegraphics[width=0.5\textwidth]{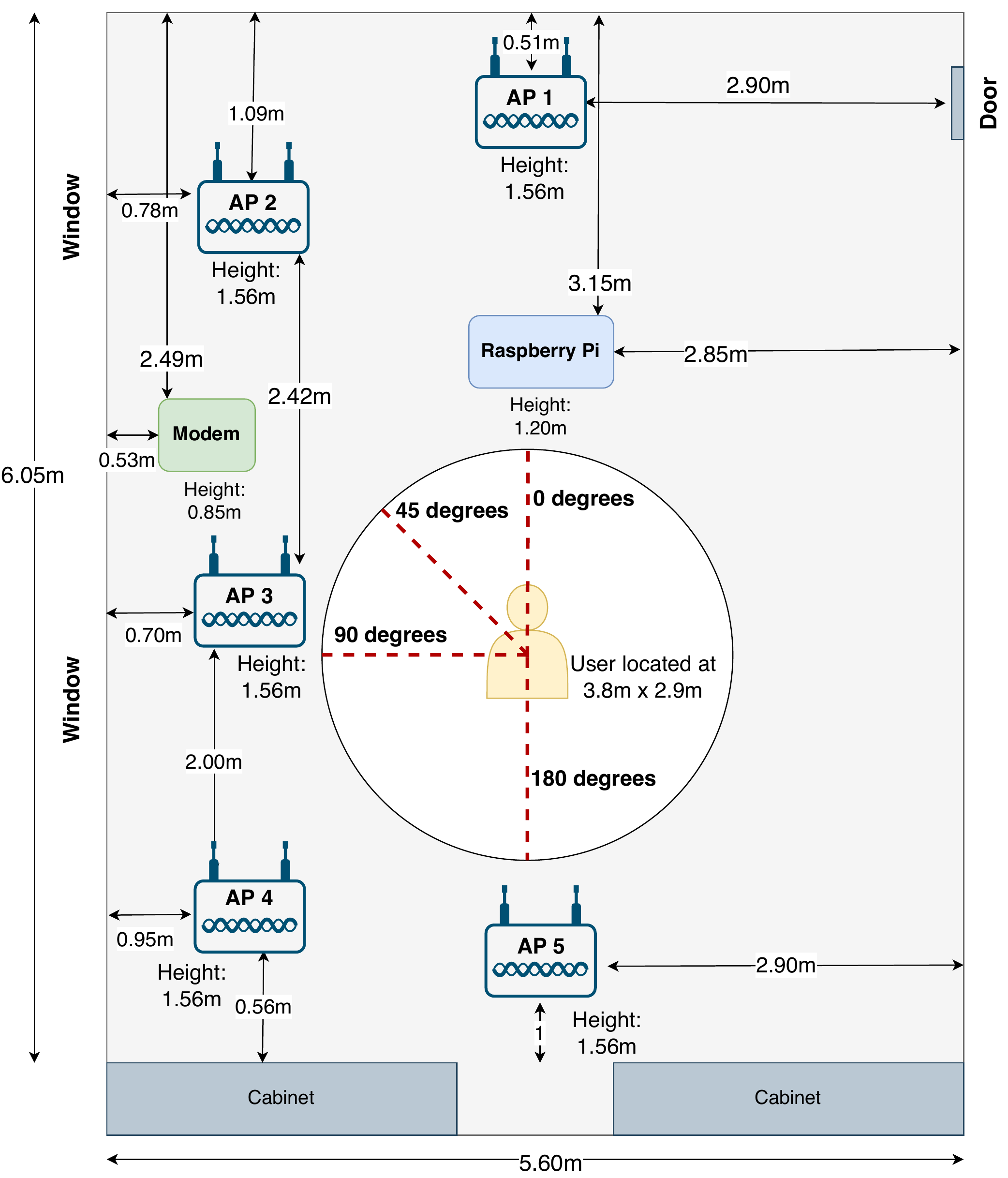} 
\caption{Detailed floor plan of the data collection setup in an indoor office, which includes location of access points (APs), user, transmitter (Raspberry Pi), and collector (modem).} 
\label{fig:floorplan}
\vspace{-4mm}
\end{figure}

\begin{table*}[]
\captionsetup{font=footnotesize}
\caption{Classification performance results and standard deviation (in $\%$) over all activity and orientation classes, averaged over $10$ independent runs. Acc:  Accuracy; BAcc: Balanced accuracy; MCC: Matthews correlation coefficient.}
\label{T:average_accuracy}
\centering
\begin{adjustbox}{width=0.8\textwidth}
\begin{tabular}{c|cccc|cccc}
\hline
\multirow{2}{*}{Model} & \multicolumn{4}{c|}{Activity}                                                                                                     & \multicolumn{4}{c}{Orientation}                                                                                                 \\ \cline{2-9} 
                       & \multicolumn{1}{c}{Acc}            & \multicolumn{1}{c}{BAcc}           & \multicolumn{1}{c}{F1-Score}       & MCC             & \multicolumn{1}{c}{Acc}            & \multicolumn{1}{c}{BAcc}           & \multicolumn{1}{c}{F1-Score}       & MCC            \\ \hline
SAP - AP 1                   & \multicolumn{1}{c}{73.3$\pm$1.8} & \multicolumn{1}{c}{73.3$\pm$1.7} & \multicolumn{1}{c}{73.3$\pm$1.8} & 64.5$\pm$2.4  & \multicolumn{1}{c}{98.0$\pm$0.5} & \multicolumn{1}{c}{98.1$\pm$0.5} & \multicolumn{1}{c}{98.0$\pm$0.5} & 97.4$\pm$0.7 \\ \hline
SAP - AP 2                   & \multicolumn{1}{c}{69.1$\pm$1.4} & \multicolumn{1}{c}{69.1$\pm$1.4} & \multicolumn{1}{c}{69.1$\pm$1.3} & 58.9$\pm$1.8 & \multicolumn{1}{c}{97.4$\pm$0.5} & \multicolumn{1}{c}{97.4$\pm$0.5} & \multicolumn{1}{c}{97.4$\pm$0.5} & 96.5$\pm$0.7 \\ \hline
SAP - AP 3                   & \multicolumn{1}{c}{70.3$\pm$3.0} & \multicolumn{1}{c}{70.3$\pm$2.9} & \multicolumn{1}{c}{70.2$\pm$3.0} & 60.4$\pm$4.0  & \multicolumn{1}{c}{98.9$\pm$0.5} & \multicolumn{1}{c}{98.8$\pm$0.5} & \multicolumn{1}{c}{98.9$\pm$0.5} & 98.5$\pm$0.7 \\ \hline
SAP - AP 4                   & \multicolumn{1}{c}{79.5$\pm$1.5} & \multicolumn{1}{c}{79.6$\pm$1.5} & \multicolumn{1}{c}{79.5$\pm$1.5} & 72.7$\pm$2.0  & \multicolumn{1}{c}{98.7$\pm$0.5} & \multicolumn{1}{c}{98.7$\pm$0.5} & \multicolumn{1}{c}{98.7$\pm$0.5} & 98.2$\pm$0.7 \\ \hline
SAP - AP 5                   & \multicolumn{1}{c}{82.7$\pm$1.5} & \multicolumn{1}{c}{82.8$\pm$1.5} & \multicolumn{1}{c}{82.7$\pm$1.5} & 77.0$\pm$2.0  & \multicolumn{1}{c}{99.4$\pm$0.4} & \multicolumn{1}{c}{99.4$\pm$0.4} & \multicolumn{1}{c}{99.4$\pm$0.4} & 99.2$\pm$0.6 \\ \hline
AMAP            & \multicolumn{1}{c}{91.1$\pm$1.8} & \multicolumn{1}{c}{91.1$\pm$1.8} & \multicolumn{1}{c}{91.1$\pm$1.8} & 88.1$\pm$2.4  & \multicolumn{1}{c}{99.0$\pm$0.1} & \multicolumn{1}{c}{99.0$\pm$0.1} & \multicolumn{1}{c}{99.0$\pm$0.1} & 99.0$\pm$0.1 \\ \hline
CMAP          & \multicolumn{1}{c}{91.4$\pm$1.4} & \multicolumn{1}{c}{91.4$\pm$1.5} & \multicolumn{1}{c}{91.4$\pm$1.5} & 88.5$\pm$1.9  & \multicolumn{1}{c}{99.7$\pm$0.2} & \multicolumn{1}{c}{99.7$\pm$0.2} & \multicolumn{1}{c}{99.7$\pm$0.2} & 99.6$\pm$0.2 \\ \hline
\end{tabular}
\end{adjustbox}
\end{table*}

\begin{table*}[]
\captionsetup{font=footnotesize}
\caption{Classification accuracy results and standard deviation (in $\%$) per activity class and orientation class, averaged over $10$ independent runs.}
\label{T:accuracy}
\centering
\begin{adjustbox}{width=0.8\textwidth}
\begin{tabular}{c|cccc|cccc}
\hline
\multirow{2}{*}{Model} & \multicolumn{4}{c|}{Activity}& \multicolumn{4}{c}{Orientation} \\ \cline{2-9} 
                       & \multicolumn{1}{c}{Circle}         & \multicolumn{1}{c}{Left-Right}     & \multicolumn{1}{c}{Push-Pull}      & Up-Down        & \multicolumn{1}{c}{$0^{\circ}$}              & \multicolumn{1}{c}{$45^{\circ}$}             & \multicolumn{1}{c}{$90^{\circ}$}             & $180^{\circ} $            \\ \hline
SAP - AP 1                   & \multicolumn{1}{c}{78.6$\pm$3.1} & \multicolumn{1}{c}{71.1$\pm$5.5} & \multicolumn{1}{c}{77.3$\pm$2.0} & 66.0$\pm$4.8 & \multicolumn{1}{c}{98.5$\pm$1.3} & \multicolumn{1}{c}{98.3$\pm$1.1} & \multicolumn{1}{c}{98.1$\pm$0.6} & 97.4$\pm$1.7 \\ \hline
SAP - AP 2                   & \multicolumn{1}{c}{72.9$\pm$3.1} & \multicolumn{1}{c}{65.7$\pm$5.6} & \multicolumn{1}{c}{71.0$\pm$3.2} & 66.9$\pm$3.3 & \multicolumn{1}{c}{96.7$\pm$1.5} & \multicolumn{1}{c}{98.0$\pm$1.1} & \multicolumn{1}{c}{98.0$\pm$1.5} & 96.8$\pm$1.5 \\ \hline
SAP - AP 3                   & \multicolumn{1}{c}{77.0$\pm$4.1} & \multicolumn{1}{c}{67.9$\pm$4.1} & \multicolumn{1}{c}{71.9$\pm$6.6} & 64.4$\pm$3.2 & \multicolumn{1}{c}{99.3$\pm$0.6} & \multicolumn{1}{c}{99.5$\pm$0.6} & \multicolumn{1}{c}{97.4$\pm$1.2} & 99.1$\pm$1.1 \\ \hline
SAP - AP 4                   & \multicolumn{1}{c}{81.3$\pm$4.2} & \multicolumn{1}{c}{78.6$\pm$2.1} & \multicolumn{1}{c}{82.5$\pm$2.5} & 75.9$\pm$5.1 & \multicolumn{1}{c}{99.5$\pm$0.7} & \multicolumn{1}{c}{98.5$\pm$1.1} & \multicolumn{1}{c}{97.7$\pm$1.2} & 99.0$\pm$1.7 \\ \hline
SAP - AP 5                   & \multicolumn{1}{c}{81.7$\pm$3.0} & \multicolumn{1}{c}{81.0$\pm$3.8} & \multicolumn{1}{c}{84.5$\pm$2.9} & 83.8$\pm$3.2 & \multicolumn{1}{c}{99.4$\pm$0.5} & \multicolumn{1}{c}{99.2$\pm$0.7} & \multicolumn{1}{c}{99.6$\pm$0.6} & 99.6$\pm$0.4 \\ \hline
AMAP            & \multicolumn{1}{c}{92.4$\pm$2.6} & \multicolumn{1}{c}{86.0$\pm$2.6} & \multicolumn{1}{c}{94.9$\pm$2.6} & 84.1$\pm$2.7 & \multicolumn{1}{c}{99.0$\pm$0.1}  & \multicolumn{1}{c}{99.0$\pm$0.1}  & \multicolumn{1}{c}{99.0$\pm$0.1}  & 99.0$\pm$0.1  \\ \hline
CMAP          & \multicolumn{1}{c}{93.6$\pm$2.4} & \multicolumn{1}{c}{89.5$\pm$3.2} & \multicolumn{1}{c}{93.4$\pm$2.6} & 88.9$\pm$3.8 & \multicolumn{1}{c}{99.6$\pm$0.7} & \multicolumn{1}{c}{99.7$\pm$0.4} & \multicolumn{1}{c}{99.9$\pm$0.3} & 99.8$\pm$0.4 \\ \hline
\end{tabular}
\end{adjustbox}
\end{table*}

\section{Experiments}
 
\subsection{Data}
We have conducted the experiments for $4$ different activity classes (\textit{Circle, Left-Right, Push-Pull, Up-Down}) as demonstrated in Figure~\ref{fig:activities}. The CSI data was collected at $4$ different orientations ($0^{\circ},45^{\circ},90^{\circ},180^{\circ}$) as demonstrated in Figure~\ref{fig:floorplan}. This figure shows our data collection setup which was conducted in an approximately $6m\times 5.6m$ indoor office with $5$ APs. The CSI of each AP was read synchronously in a central collector. A Raspberry Pi was used as the transmitter. Per each combination of orientation class and activity class, $20$ samples were collected from $6$ users. The total number of collected samples from each AP was $20\times 4\times 4\times 6=1,920$, where $80\%$ was used for training and $20\%$ was used for testing the models. The dataset will become publicly available for the research community.

\subsection{Setup}
The Ridge regression classifiers were cross-validated with $(0.001,0.01,0.1,1)$ regularization strengths. The reported results are averaged over $10$ independent runs. We have partially used the PyTorch implementation\footnote{https://github.com/timeseriesAI/tsai/blob/main/tsai} of the MiniRocket~\cite{dempster2021minirocket} with a fixed set of $K=84$ kernels of length $9$ and the total number of features of $D=9,996$. Our codes are available online\footnote{https://github.com/salehinejad/CSI-joint-activ-orient}. The models are implemented in PyTorch and were trained on a single NVIDIA GTX GPU.

\subsection{Classification Performance Analysis}
Classification performance of the SAP, AMAP, and CMAP models with respect to the accuracy (Acc), balanced accuracy (BAcc), F1-Score, and Matthews correlation coefficient (MCC) metrics is presented in Tables~\ref{T:average_accuracy} and~\ref{T:accuracy}. 
 
In Table~\ref{T:average_accuracy}, the average performance results over all activity and orientation classes are presented. The SAP model was trained and evaluated per each AP independently. The results show that the SAP model with AP $5$ has a better performance than the other SAP models for both activity and orientation recognition tasks. As Figure~\ref{fig:floorplan} shows, the user is located between the shortest path between the AP and the transmitter. However, the other APs have a shortest LoS with the transmitter without direct interference with the user. Hence, proper placement of the APs with respect to the user and transmitter location can improve sensing of the environment and achieving a higher activity and orientation recognition accuracy.   
 
The overall results show that the CMAP and AMAP approaches have a competitive performance, better than the SAP evaluations. CMAP performs slightly better than AMAP in activity recognition but has a lower performance in orientation prediction. All approaches, even with a single AP, have a high performance in prediction of the orientation of the user. This is particularly important in recognizing which device/orientation a user is interacting with.

Granular performance results per activity class and orientation class in Table~\ref{T:accuracy} show that the \textit{Up-Down} activity is relatively more challenging to recognize than the other gestures. The performance per orientation class is high for all approaches and the performance difference between different classes is not significant. Overall, the CMAP approach has a relatively better performance and less complexity due to using a single joint activity and orientation classifier.

\section{Conclusions}
\label{sec:conclusion}
In this paper, for the first time in the literature, we explore joint prediction of human's orientation and activity using WiFi signals for human-machine interaction in indoor environments. In order to be able to deploy the solutions on resource-limited devices, models based on random convolution kernels without training them are proposed for feature extraction. The simple but effective Ridge regression classifier is used for features classification. Our results show that increasing the spatial diversity of WiFi signal collection by utilizing multiple APs can increase the classification accuracy of human activities. However, it is possible to predict orientation of the user using a single AP with a high accuracy.

% \section{Acknowledgment}
% This work was partially supported by the Mobile AI Lab established between Huawei Technologies Co. LTD Canada
% and The Governing Council of the University of Toronto.
\bibliographystyle{IEEEbib}
\bibliography{strings,mybibfile}

\end{document}